\documentstyle[12pt]{article}

\def\frac#1#2{{\textstyle {#1 \over #2}}}

\def\Eq{\begin{equation}}   \def\Endeq#1{\label{#1} \end{equation}}
\def\Eqa{\begin{eqnarray}}  \def\Endeqa#1{\label{#1} \end{eqnarray}}

\newcommand{\Eqr}[1]{(\ref{#1})}

\def\&{and}

\def\DS {D\!\!\!\!/}
\def\EA {\Gamma [ A_{\mu} ]}
\def\np#1#2#3{           {\it Nucl. Phys. }{\bf #1}, #2 (19#3)}
\def\pl#1#2#3{           {\it Phys. Lett. }{\bf #1}, #2 (19#3)}
\def\pr#1#2#3{           {\it Phys. Rev. }{\bf #1}, #2 (19#3)}
\def\prep#1#2#3{         {\it Phys. Rep. }{\bf #1}, #2 (19#3)}

\begin{document}
\begin{titlepage}

\begin{center}
September, 1992      \hfill       HUTP-92-A048\\
\vskip .5 in
{\large \bf A New Anomaly Matching Condition?}
\vskip .3 in
{
  {\bf  Vineer Bhansali\footnote{Email: \tt Bhansali@HUHEPL.bitnet} and
Stephen D.H. Hsu}\footnote{Junior Fellow,
Harvard Society of
     Fellows. Email: \tt Hsu@HUHEPL.bitnet, Hsu@HSUNEXT.Harvard.edu}
   \vskip 0.3 cm
   {\it Lyman Laboratory of Physics,
        Harvard University,
        Cambridge, MA 02138}\\ }
  \vskip 0.3 cm
\end{center}

\vskip .5 in
\begin{abstract}
We formulate ``Witten'' matching conditions for confining gauge
theories.
The conditions are analogous to 't Hooft's, but involve Witten's
global SU(2) anomaly. Using a group theoretic result of Geng,
Marshak, Zhao and Okubo, we show that if the fourth homotopy group of
the flavor group $H$ is trivial ($\Pi_4(H) = 0$) then realizations of
massless composite fermions that satisfy the
't Hooft conditions also satisfy the Witten conditions. If $\Pi_4 (H)$
is nontrivial, the new matching conditions can yield additional
information
about the low energy spectrum of the theory. We give a simple
physical proof of Geng, et. al.'s result.

\end{abstract}
\end{titlepage}

\renewcommand{\thepage}{\arabic{page}}
\setcounter{page}{1}
\section{The Global SU(2) Anomaly }
\renewcommand{\thepage}{\arabic{page}}
Witten \cite{WGA} has shown that SU(2) gauge theories are subject to
a nonperturbative anomaly which can render them mathematically
inconsistent.
For example, SU(2) theories with an odd number of chiral doublets
(and no other fermions in higher representations) are inconsistent. The
reason behind the inconsistency is as follows: The fourth homotopy
group of SU(2) is nontrivial, $\Pi_4( SU(2) ) = Z_2$. This means that
there is a homotopically nontrivial class of four dimensional SU(2)
gauge configurations that cannot be continuously deformed to the
identity. Now consider the fermion integration in the Euclidean path
integral for an odd number $N$ of Weyl fermions:
\begin{equation}
\int D\psi D\bar{\psi} ~exp(\int d^4x~ \sum_j^N\bar{\psi}_j i \DS
\psi_j)
   = {\det}^{N/2}[ i \DS(A) ],
\end{equation}
where A is the background gauge field. It can be shown using the
Atiyah-Singer index theorem \cite{AS} that the fermion determinant,
defined as the product of either the positive or negative eigenvalues
of $i\DS(A)$, changes sign under a topologically nontrivial gauge
transformation U:
\begin{equation}
{\det}^{N/2}[ i \DS(A) ] = (-)^N~ {\det}^{N/2}[ i \DS(A^U) ] ,
\end{equation}
where $A^U_{\mu} = U^{-1}A_{\mu}U - iU^{-1} \partial_{\mu} U$. Hence
for $N$ odd, when the integration over gauge configurations $A_{\mu}$
is performed in the partition function Z, the homotopically trivial
and nontrivial gauge sectors cancel exactly, yielding zero.
Similarly, the path integral $Z_X$ with insertion of any gauge
invariant operator X is identically zero. Therefore any expectation
value $\langle X \rangle = Z_X / Z$ is ill defined.

For the more general case where there exist not only fermion doublets but also
other representations, $N$ in the above equations gets replaced by $N_0$,
which is the number of zero modes in an instanton background gauge field and
equals $N_0 = \sum_{R_i} 2 T(R_i)$ where $R_i$ is an $SU(2)$ representation
and $T(R_i)$ is the index defined by ${\rm Tr} \, T_aT_b|_{R_i} = T(R_i)
\delta_{ab}$.

The Witten anomaly, unlike the familiar triangle anomaly, is
intrinsically nonperturbative and is often referred to as a {\it
global } anomaly (this is not to be confused with anomalies in global
currents, which we will not discuss here). While the triangle
anomaly, due to its non-renormalization
properties, can be computed at one loop order, the Witten anomaly
does not appear at any order in perturbation theory.

In this letter
we will discuss some issues related to the Witten anomaly. The first
is its use in deriving ``Witten'' matching conditions, similar to
those of 't Hooft \cite{TH}, which constrain the possible
realizations of massless composite fermions in confining gauge
theories. Such matching conditions are remarkable as they yield
nonperturbative, dynamical information in terms of simple algebraic
relations.
We will find, perhaps suprisingly since the two types of anomaly are
seemingly very different, that if the full flavor group $H$ satisfies
$\Pi_4(H) = 0$ then a realization of massless composite fermions
which satisfies the 't Hooft matching conditions automatically
satisfies the Witten matching conditions. This result follows from
the group theoretic results of Geng, Marshak, Zhao and Okubo
regarding global and local anomaly cancellation.
We will find that their results can be neatly understood in a
physical way in terms of the gauge invariance properties of gauge
field effective actions.
Alternatively, if $\Pi_4(H)$ is nontrivial, the Witten matching
conditions can yield new information about the massless bound states in
the theory.

\section{Anomaly Matching Conditions}
Consider a confining gauge theory with color group G and fundamental
chiral fermion fields $\psi$. Let the group of global flavor
symmetries for this theory be $H$.
If the global chiral symmetries
are not broken in the confined phase of this theory, the physical
gauge singlet states will include  massless composite fermions
(``baryons'' - denoted here by B).
The B's will form complicated
representations of $H$, as typically many $\psi$'s will have to be
combined to a yield gauge singlet state. 't Hooft's matching conditions
\cite{TH}
provide very nontrivial constraints on the possible representations
B.
To arrive at his conditions 't Hooft considered gauging the flavor
group with an extremely small gauge coupling (so as not to affect the
strong color dynamics). In order to have a consistent theory, 't Hooft
introduced massless spectator fermions S which carry flavor
charges, but are color singlets. The spectators are chosen to exactly
cancel any triangle anomalies due to the fundamental fermions $\psi$.
Symbolically,
\begin{equation}
A^{\triangle} _{\psi} = - A^{\triangle}_S.
\end{equation}
Now consider the low energy, confined limit of the theory. One
expects the effective theory here to be described by the massless
composites, which now carry only the flavor charge. However, a
consistent effective theory requires that the triangle anomalies of
the composite fermions cancel those of the massless spectators.
Therefore, one has the relation
\begin{equation}
A^{\triangle}_B = - A^{\triangle}_S = A^{\triangle}_{\psi}.
\end{equation}
In other words, the perturbative flavor anomalies of the composite
fermions must exactly equal those of the fundamental fermions. Since
the possible representations B are limited by the requirement of
color neutrality, these
matching conditions are in many cases nontrivial to satisfy. If they
cannot be satisfied one of the assumptions of the construction must
be abandoned - for example, conservation of chiral symmetries or
confinement. If one assumes confinement, then this line of reasoning,
combined with the decoupling or ``persistence of mass'' \cite{PW} condition,
shows that QCD must break its chiral symmetries $SU(n_f)_L \otimes SU(n_f)_r
\otimes U(1)_V$
down to $SU(n_f)_V \otimes U(1)_V$, if $n_f > 2$. Persistence of mass, which
requires that a {\it massless} composite not contain a {\it massive}
constituent, has been demonstrated rigorously for vectorlike theories by Vafa
and Witten \cite{VW}. (See \cite{HSU} for a generalization of this result when
fundamental scalars are present.)

More detailed and rigorous arguments for 't Hooft's conditions have
been given by Frishman, Schwimmer, Banks and Yankielowicz \cite{FSBY}
and by Coleman and Grossman \cite{CG}. Their strategy is to compare
the anomalies of the fundamental and bound state spectra by comparing
 the singularities
and discontinuities of the anomaly equation implied by the triangle diagram
in the confined and deconfined phases of the theory.

We can now generalize 't Hooft's construction by considering SU(2)
subgroups of $H$, and adding spectator fermions $S'$ (note that $S'$
and S are not {\it a priori} the same representations) to cancel the
global anomaly of the fundamental fermions, $\psi$. Reasoning
similar to that of the above analysis yields the condition:
\begin{equation}
A^W_B = A^W_{\psi}.
\end{equation}
In other words, the composite fermions must exactly reproduce the
Witten anomaly of the fundamental fermions. Of course, the global
matching conditions cannot be justified by examining the structure of
specific perturbative graphs such as the triangle, and so the above
result cannot be justified at the same level of rigor as the 't Hooft
conditions.
However, if correct, as the spectator argument suggests,
the above conditions {\em seem} to again yield, at first sight,
nontrivial constraints on
the representations B, and can provide new information about the
bound state spectrum of chiral gauge models.

Unfortunately, in general this is not the case. It is possible to show
that a set of massless composite fermions B which satisfy the
't Hooft conditions (4) will also {\it automatically} satisfy the
Witten matching conditions (5) if $\Pi_4(H) = 0$. In order to prove
this, we will invoke the result of Geng, Marshak, Zhao and Okubo
(GMZO) \cite{GMZO}, which states that a theory with simple gauge
group H, which satisfies $\Pi_4 (H) = 0$ and which has no
perturbative anomalies, will have no global anomalies in any of its
SU(2) subgroups. The GMZO result is entirely group theoretical, and
relates the global SU(2) anomaly to the local anomalies of
the larger group H.  It is important to emphasize that the
GMZO result states that the vanishing of the
triangle anomaly implies the vanishing of the Witten anomaly,
but not vice versa. (A model can still have
vanishing Witten anomaly and non-vanishing triangle anomaly.)
In the next section we will explain the GMZO result in more detail, from
a more physical viewpoint.

Now suppose that the composite fermions B satisfy the 't Hooft
conditions (4). Then both the short distance, fundamental theory and
the low energy, composite theory are free from perturbative anomalies when
the spectators S are included. If $\Pi_4 (H) = 0$ and
$H$ is simple then the GMZO result applies, and we can conclude that
no SU(2) subgroup of $H$ has a Witten anomaly when the spectators are
included. But this just implies that
\begin{equation}
A^W_B = - A^W_S = A^W_{\psi} \label{condition} ,
\end{equation}
which is the Witten matching condition.  Here the first equality uses GMZO,
and the second equality uses the assumption of unbroken $H$, 't Hooft's
matching condition, and GMZO.
The condition \Eqr{condition} also implies that the
spectators S from the 't Hooft construction will {\em suffice} as the
spectators $S'$ for the corresponding Witten construction.
(This is always the case for any model, as we can always choose the spectators
to be ``mirror'' fermions, whose addition renders the model vectorlike.)
  But note that
in principle it is possible to have Witten anomaly cancellation without having
't Hooft anomaly cancellation.

Now let us consider the possibility that $\Pi_4(H)$ is nontrivial. In that
case the GMZO result cannot be directly applied. For the simplest
case, $H = SU(2)$, it is easy to see that the 't Hooft conditions are
trivially satisfied due to the fact that representations of SU(2) are
pseudo-real and hence cannot contribute to the triangle anomaly.
Therefore, if the Witten conditions are not always trivially
satisfied, they provide new constraints on the spectrum of
massless composites.  Here we will
consider two specific models: an $SU(N_c) \otimes SU(2_f)_L$,
and an $SU(3_c) \otimes SU(2_f)_L \otimes SU(2_f)_R$, where the subscripts $c$
and $f$ denote color and flavor respectively.

1. $SU(N_c) \otimes SU(2_f)_L$:  We take $N_c$ colors of fundamental fermions
$X$
transforming as doublets ($2$'s) of $SU(2_f)$. In order to cancel color
anomalies, it will be necessary to include additional left and right handed
fermions in possibly higher color representations. As a specific example,
for $N_c = 3$,
consider adding a left handed flavor singlet which transforms as a
\bf 15 \rm of color, $Y$, and 16 right handed flavor singlets which are
triplets of
color, $Z$. The net color anomaly \cite{BG}
(in units of the fundamental 3) is then \footnote{The interested reader can
construct more sophisticated models along these lines by using the toolbox
for computing indices given in the appendix.}
$A = 2 + 14 -16 = 0$, where the first contribution is from each member of
the flavor doublet, the second one from the \bf 15 \rm of color (a flavor
singlet), and the final contribution is from the 16 right handed (hence the
relative minus sign) color
triplets (which are flavor singlets). The full symmetry of the model is
$SU(3_c) \otimes SU(2_f)_L \otimes SU(16_f)_R \otimes U(1)^3$,
where the $U(1)$'s are associated with phase rotations of the
three types of fermion. One problem with this specific model
is that the addition of the extra fermions destroys
asymptotic freedom. This may not be a problem for
similar constructions with larger $N_c$. In any
case, the group theory still provides an example of new
information from
Witten matching conditions.

In this model there are Lorentz invariant color singlet operators which
transform under $SU(16_f)_R \otimes U(1)^3$, but are singlets under
$SU(2_f)_L$. For instance,
\begin{equation}
( \epsilon^{ij} X^{\alpha}_i  X^{\beta}_j )^3
\end{equation}
carries the first $U(1)$ charge, but is an $SU(2_f)_L$ singlet
($i,j$ are flavor indices and $\alpha,\beta$ are color indices).
We can also form operators out of $Y$ and $Z$ fields that are
spin zero, color and $SU(2_f)_L$ singlet, but transform under
$SU(16)$ and the remaining two $U(1)$'s.
We can consider the case where condensates of the above operators form,
breaking all the chiral
symmetries except $SU(2_f)_L$.
Then the 't Hooft anomaly
matching condition is trivial, since there is no triangle anomaly for the
$SU(2)$ group because all representations are pseudoreal.
Similar constructions are possible for larger values of $N_c$.

In models of this sort
it is
possible to obtain nontrivial constraints by considering the Witten anomaly
matching condition presented above.
To classify the color singlet baryons in the confined theory, we
consider the tensor product $\underbrace{{\bf 2} \otimes {\bf 2} \otimes \ldots
{\bf 2}}_{N_c \,
{\rm times} \,}$,  and decompose this into representations of the flavor
group $SU(2_f)$.
GMZO point out that there are an even number of
fermion zero modes in any representation of the $SU(2)$ gauge group
except those of dimension
\Eq
n=2(2m+1), \, m = 0,1,2,\ldots
\Endeq{zero}
The number of zero modes is given by $N_0 = {1 \over 6} n (n^2-1), \,
n=1,2,3,\ldots$ which is odd only for $n$ given in \Eqr{zero}.
Therefore it is only representations of
dimensionality $n$ that contribute to the Witten anomaly.
These `anomalous' $n$ are all even, i.e. $n = 2,6,10,14,18,\ldots$

It is easy to see from Young tableaux
that only two kinds of irreps may
appear in the decomposition of
the product of $N_c$ {\bf 2}'s, all even-dimensional or all odd dimensional.
When $N_c$ is even, the only irreps are of dimensions $N_c+1, N_c-1, N_c-3,
\ldots, 1$, i.e. all odd dimensional. Since \Eqr{zero} requires even
dimensional irreps. for a Witten anomaly, it is clear that when $N_c$ is even,
it is impossible to get a Witten anomaly in the confining theory.  This is
consistent, since for even $N_c$, the fundamental theory does not have a Witten
anomaly either.  In the case $N_c$ even, the Witten anomaly matching thus
gives no
new constraints on the representations of the flavor group in the confining
phase.

For $N_c$ odd, we know that the fundamental theory has a Witten anomaly
which must be reproduced in the low energy confining theory.  For an $N_c$-
fold tensor product of the $2$'s transforming under $SU(2_f)$, ($N_c$ odd)
it is again easy to see that the only irreps appearing are of dimension
$N_c+1, N_c-1, N_c-3, \ldots, 0$, i.e. all even dimensional.  Thus for odd
$N_c$, there must be at least one massless `baryon' that transforms under
$SU(2_f)$ as one of the anomalous representations given by \Eqr{zero}.
The Witten matching condition requires irreps that replicate the
Witten anomaly in the low energy theory, i.e. the ones of dimension
$2,6,10,\ldots$. Massless composites must appear in at least one of
those irreps, a fact which could not have been deduced from the 't Hooft
conditions.

2. $SU(3_c) \otimes SU(2_f)_L \otimes SU(2_f)_R \otimes U(1)_V$:
Since this is a vector-like theory, the perturbative anomaly constraints can
be combined with the persistence of mass condition \cite{PW,VW}. However,
there are many solutions to these combined conditions and it is therefore
not possible to prove that chiral symmetry breaking occurs for two flavors.
(It may be that a light strange quark is necessary for chiral symmetry
breaking in QCD!)

Perhaps the Witten conditions can eliminate all or some of the solutions:
consider gauging the $SU(2_f)_L$ gauge group. For $N_c =3$, the 't Hooft
matching
equation obtained from the $[SU(2_f)_L]^2 \otimes U(1)_V$ matching
condition is then \cite{TH}
\Eq
10 a - 5 b + c =1
\Endeq{acth}
where $a,b$ and $c$ are non-negative integers denoting
respectively the number of fermions transforming under $SU(2_f)_L \otimes
SU(2_f)_R$ as $({\bf 4},{\bf 1})$, $({\bf 2},{\bf 3})$ and $({\bf 2},{\bf
1})$.  The 't Hooft anomaly condition is satisfied by any set of $a,b,c$
which satisfies \Eqr{acth}.
There is also an $SU(2)_L$ Witten anomaly in the
fundamental theory with three colors,
which needs to be matched in the confining theory.

As a special case, choose $a=0, b=0, c=1$ in the 't Hooft matching equation
which gives the low energy spectrum containing simply the representation
$({\bf
2},{\bf 1})$ and its parity double $({\bf 1},{\bf 2})$, which are the nucleons
of the $\sigma$ model. The representation $({\bf 2},{\bf 1})$ is also one of
those in \Eqr{zero} that can give a Witten anomaly, so the Witten anomaly is
also satisfied.

Another solution of the 't Hooft matching equation is afforded by $a=1, b=2,
c=1$, which corresponds to a low
energy spectrum of $({\bf 4},{\bf 1}) \oplus 2({\bf 2},{\bf 3}) \oplus
({\bf 2},{\bf 1})$ plus their parity doubles.  The 't Hooft
conditions cannot distinguish this realization from the previous one.
Since by \Eqr{zero} the $({\bf 4},{\bf 1})$ representation gives no
contribution
to the Witten anomaly, we need only consider the contribution of the
irreps $({\bf 2},{\bf 3})$
and $({\bf 2},{\bf 1})$.  In toto, these two
 contribute seven $SU(2_f)_L$ doublets,
so the Witten matching condition is again
satisfied, yielding no new information.

In general, only
the irreps $({\bf 2},{\bf
3})$ or $({\bf 2},{\bf 1})$ can contribute to the
Witten anomaly in the confined phase, so we can ignore the index $a$.
If the Witten matching conditions are to eliminate
representations allowed by
the 't Hooft matching conditions, there must exist
$b$ and $c$ such that $3b+c$ is
even.  Setting $3b+c=2m, \, m = 0,1,2,\ldots$,
and using the 't Hooft equation, we obtain $10a - 8b +2m = 1$, which can never
 be
satisfied by integral $a,b$ and $m$.  In this case we conclude that the
Witten anomaly matching conditions are subsumed by the 't Hooft anomaly
matching conditions. We have checked that this is the also the case in
five color QCD with two flavors, with or without parity doubling of nucleons.

It is possible to give a general proof of this result for any
$SU(N_c) \otimes SU(2_f)_L \otimes SU(2_f)_R \otimes U(1)_V$ theory.
($N_c$ must be odd in order that the confined phase have chiral baryons with
 spin $1/2$). Consider the 't Hooft matching condition resulting from the
 $[ ( SU(2_f)_L )^2 U(1)_V ]$ anomaly. In the fundamental theory we have
 $N_c$ left handed doublets of $SU(2_f)_L$, which contribute $N_c/2$ to the
 mixed anomaly (we define $T(R)=1/2$ in the fundamental representation, where
 $T(R)$ satisfies $Tr T^a T^b|_R = T(R) \delta^{ab}$).
In the confined phase of the theory we can have any number of baryons
transforming in higher representations of $SU(2_f)_L$. However, in the
simplest case these baryons will consist of $N_c$ fundamental fermions,
and therefore have $U(1)_V$ charge $N_c$. The anomaly matching equation is then
\begin{equation}
\sum_{i} 2 l_i ~T(R_i) = 1,
\Endeq{tmc}
where the sum is over all baryon states, with multiplicity $l_i$. Now consider
Witten matching conditions, which require an odd number $N_0$ of zero modes in
 an $SU(2_f)_L$ instanton background (recall $N_c$ is odd, so we have an odd
 number of doublets in the fundamental theory).
 Since  $N_0 (R_i) = 2 T(R_i)$  \cite{GMZO}, the Witten condition becomes
 \begin{equation}
\sum_i  2 l_i ~T(R_i) = \mit{odd}.
\Endeq{wmc}
It is clear that \Eqr{wmc} is implied by
\Eqr{tmc}. Note that if the $U(1)$ symmetry is
spontaneously broken in the low energy theory (as was assumed in
the first example), \Eqr{tmc} no longer applies
and the Witten conditions may contain new information.

\renewcommand{\thepage}{\arabic{page}}
\section{Perturbative and Global Anomalies}
In this section we address the relation between
perturbative and global anomalies. This relation
will yield a particularly simple understanding of the
GMZO result. The point of view taken here follows that of
Alvarez-Gaume and Witten \cite{AW}.

Consider integrating out chiral fermions in the
background of an arbitrary gauge field $A_{\mu}$
(see equation (1)). The result is an effective action
\begin{equation}
\Gamma [ A_{\mu} ] = - \ln~ {\det}^{N/2}[ i \DS(A_{\mu}) ].
\end{equation}
If we wish to quantize the gauge fields
(i.e.  to treat $A_{\mu}$ as dynamical quantum fields),
we require that the effective action
$\Gamma [ A_{\mu} ]$ be invariant under
all gauge transformations. Gauge invariance is
a classical symmetry of the theory, but may be
violated at the quantum level by the introduction of chiral fermions.
When $\Gamma [ A_{\mu} ]$ is not gauge invariant, we say that the theory
has a gauge anomaly.

Perturbative, or triangle, anomalies
correspond to noninvariance of $\EA$
under gauge transformations $U$ which
can be smoothly deformed to the identity.
The explicit form of the effective action
was computed by Wess and Zumino \cite{WZ}.
Global anomalies correspond to noninvariance of
$\EA$ under $U$'s which are nontrivial mappings of
$S^4$ into the gauge group. In a theory with a global
anomaly the fermion determinant changes sign under such
a nontrivial gauge transformation $U_*$. But this tells
us that in such a theory
\begin{equation}
\Gamma [ A^{U_*}_{\mu} ] = \Gamma [ A_{\mu} ] +  i m \pi,
\end{equation}
where $m$ is odd. This was originally noted in \cite{WCA}.

Now we turn to the GMZO result, which can be easily formulated in this
language.
Consider a theory with gauge group $H$ and chiral fermions $\psi$.
Suppose that this theory exhibits neither a Witten anomaly (for
example, $\Pi_4(H) = 0$) nor a triangle anomaly. Then $\EA$ is
completely gauge invariant and we can define a consistent quantum
gauge theory based on $H$ with fermions $\psi$.
Now consider any $SU(2)$ subgroup of $H$ (or in
general any subgroup $H'$ with nontrivial $\Pi_4$).
It is clear that $H'$ cannot have either a global or
local anomaly. Either type of anomaly would require noninvariance of
$\Gamma [ A^{H/H'}_{\mu} = 0, A^{H'}_{\mu} ]$, where $A^{H'}$
are gauge fields in $H'$ and $A^{H/H'}$ are gauge fields in $H/H'$.
$\Gamma [ A^{H/H'}_{\mu} = 0, A^{H'}_{\mu} ]$ is simply the effective action
for gauge fields of $H'$ which arises from integrating out the fermions
$\psi$. However, since by assumption $\Gamma [ A^{H/H'}_{\mu}, A^{H'}_{\mu} ]$
is completely gauge invariant, $H'$ cannot have either a global or local
anomaly.

In other words, the
nontrivial gauge transforms $U_* \subset H'$ are a subset of the
gauge transformations of the full theory, $U \subset H$. Indeed, if
$\Pi_4(H) = 0$ then the $U_*$ maps can be continuously deformed to
the identity in $H$. Therefore, an anomalous global transformation
in $H'$ must correspond to an anomalous local transformation in $H$.
If these are absent from the full theory (cancelling triangle anomalies
for H), then they must be absent from $H'$ (no global anomaly in $H'$).

Note that these results are somewhat more general those of GMZO. They
apply to groups $H$ which are not simple, and also to groups $H$ with
$\Pi_4(H) \neq 0$ but in which the Witten anomaly is cancelled.

Finally, we mention an even more physical argument\footnote{SDH
thanks Richard Holman for discussions on this subject.}
 for the GMZO result. Suppose, contrary to GMZO, that
one could formulate a theory with no triangle or Witten
anomalies, but which has a Witten anomaly in
  a subgroup $H'$. Then, one could consider
adding scalars whose vacuum expectations break
the full group to $H'$ while not affecting the
fermions. As the vacuum expectation values $v$
are taken to infinity the gauge fields
corresponding to broken generators become arbitrarily massive. We are then
left with a trivial effective low energy theory - any correlator computed to
lowest order in $1/v$ is exactly zero! This is extremely pathological behavior
for a theory which is perfectly well defined at short (less than
$v^{-1}$) distances. Surely nature cannot allow the realization of such a
model.

\section{Acknowledgments}

The authors would like to thank Howard Georgi, Richard Holman and Shane
Hughes for useful discussions.
 VB and SDH acknowledge
support from the National Science Foundation under grant
NSF-PHY-87-14654,
the state of Texas under grant TNRLC-RGFY106 and from the White Horse
and Railroad Foundation. SDH also acknowledges support from the
Milton Fund of the Harvard Medical School and from the Harvard
Society of Fellows.

\appendix

\section{A Group Theory Toolbox}

Here we collect some important
group theoretical results that are utilized in this paper.
For any SU(N), and indeed for any classical Lie algebra H, a famous theorem
due to Dynkin states that a sequence of rank(H) non-negative integers is a
highest weight vector for some unique irreducible representation. The
converse is also true, that
any irreducible representation can be constructed by specifying a suitable
sequence of non-negative integers from which the whole irrep.\ can be
constructed by the action of the lowering operators.

Specify a highest weight for SU(N) (rank N-1) by N-1 non-negative
integers
\begin{equation}
\Lambda \equiv (a_1,a_2,\ldots,a_{N-1}).
\end{equation}
In terms of Young tableaux, this is
just the tableaux which has $a_i$ boxes over-hanging between row $i$ and
row $i+1$. Every Young tableaux contructed by this prescription is
admissible and corresponds to a valid SU(N) tensor. The dimension of this
irrep.\ can then be computed using the factors over hooks rule.
For example, in SU(3), the Dynkin labels $(1,0)$, $(0,1)$ and $(1,1)$
correspond to the ${\bf 3}$, ${\bf {\bar{3}}}$ and ${\bf 8}$ respectively.

The advantage of using the Dynkin language is that not only can all irreps.\
be constructed by subtracting the rows of the Cartan matrix from the
highest weight (the number of
times indicated by the Dynkin indices), but also that it is rather
straightforward to compute the eigenvalues of the invariant Casimir
operators of both second and third order, which have appeared in this paper
a number of times: the quadratic index has appeared in the computation of
the number of zero modes and in the computation of the mixed $SU(2)^2
\otimes U(1)$ anomaly, whereas the cubic index has appeared in the
computation of the pure $SU(2)^3$ anomaly contribution.

For SU(3), we specify the highest weight of an irrep.\ by $(a_1,a_2)$.
Then the dimension is
$N(a_1,a_2) = (a_1+1)(a_2+1)({a_1+a_2\over 2}+ 1)$, the quadratic index is
$Q_2(a_1,a_2)={N(a_1,a_2)\over 12} (a_1^2+3a_1+a_1a_2+3a_2+a_2^2)$, and the
anomaly index is $Q_3(a_1,a_2)={N(a_1,a_2) \over 60}
(a_1-a_2)(a_1+2a_2+3)(2a_1+a_2+3)$ \cite{BG}.  A real representation of any
SU(N) is
obtained if reversing the order of the Dynkin indices leaves the irrep.\
unchanged.  For SU(3) irreps., if $a_1 = a_2$ we hence obtain a real
irrep.\ , and since $Q_3$ is then zero, we find that the real representations
(e.g. the adjoint), do not contribute to the anomaly.
It is not true though that only real irreps. are anomaly free for higher
SU(N) groups.  For instance, the 3048474 dimensional
Dynkin irrep.\ $(5,1,8,1)$ is complex but
anomaly free in SU(5).
Note also that on complex conjugation of irreps., i.e. $\Lambda
\leftrightarrow \bar{\Lambda}$, $Q_2(\Lambda) = Q_2(\bar{\Lambda})$, but
$Q_3(\Lambda) = -Q_3(\bar{\Lambda})$, which is relevant when the
contribution of opposite handedness fermions is included.

The total number of zero modes in an instanton background field is given by
the SU(2) quadratic index.  In the normalization of this paper, for an
isospin $I$ representation, it is given by $N_0 = {2\over 3}I(I+1)(2I+1)$.
This same index, under a different guise,
 also appears in the computation of the mixed anomaly
matching condition of the examples given in this paper.
For tables of the quadratic indices for most Lie algebras of interest, and
for other group theoretical data, please refer to \cite{SL}.

\vskip 1 in
\end{document}